\documentclass[prl,twocolumn,showpacs,superscriptaddress]{revtex4}
\usepackage[dvips]{graphicx}
\usepackage{amsmath,amsfonts,amssymb}
\usepackage{dcolumn}
\usepackage{bm}
\usepackage{epsfig}

\begin{document}
\newcommand{\beq}{\begin{equation}}
\newcommand{\eeq}{\end{equation}}
\newcommand{\vecE}{{\bf E}}
\newcommand{\vecF}{{\bf F}}
\newcommand{\vecv}{{\bf v}}
\newcommand{\vecA}{{\bf A}}
\newcommand{\vecp}{{\bf p}}
\newcommand{\vecn}{{\bf n}}
\newcommand{\vecr}{{\bf r}}
\newcommand{\vektr}{{\bf r}}
\newcommand{\diff}{\mathrm{d}}
\newcommand{\imagi}{\mathrm{i}}
\newcommand{\vecR}{{\bf R}}
\newcommand{\del}{\mbox{\boldmath{$\nabla$}}}
\title{Strong Field Ionization Rate for Arbitrary Laser Frequencies}

\author{S.V.\ Popruzhenko}
\affiliation{Max-Planck-Institut f\"ur Kernphysik, Postfach 103980, 69029 Heidelberg, Germany}
\affiliation{Moscow State Engineering Physics Institute, Kashirskoe Shosse 31, 115409, Moscow, Russia}
\author{V.D.\ Mur}
\affiliation{Moscow State Engineering Physics Institute, Kashirskoe Shosse 31, 115409, Moscow, Russia}
\author{V.S.\ Popov}
\affiliation{Institute for Theoretical and Experimental Physics, Bolshaya Cheremushkinskaya 25, 117218, Moscow, Russia}
\author{D.\ Bauer}
\affiliation{Max-Planck-Institut f\"ur Kernphysik, Postfach 103980, 69029 Heidelberg, Germany}

\date{\today}

\begin{abstract}
A simple, analytical, nonrelativistic ionization rate formula for atoms and positive ions in intense ultraviolet and x-ray electromagnetic fields  is derived. The rate is valid at arbitrary values of the Keldysh parameter and confirmed by results from {\em ab initio} numerical solutions of the single active electron, time-dependent Schr\"odinger equation. The proposed rate is particularly relevant for experiments employing the new free electron laser (FEL) sources.
\end{abstract}
\maketitle

In the year 2000, when a free electron laser (FEL) came into operation at DESY \cite{FEL0}, a new era of strong field laser physics has begun: the interaction of intense, ultraviolet and x-ray laser pulses with matter. 
Presently, intensities $\simeq 10^{16}$\,W/cm$^2$ are available at photon energies $\hbar\omega\simeq 100$\,eV \cite{sorokin}.
One of the most fundamental quantum processes in intense laser-matter interaction is the nonlinear photoeffect where several photons are absorbed by the emitted electron.
Ionization of atomic targets by intense pulses from FELs has been studied in several experiments \cite{wabnitz02,laarmann04,wabnitz05,mosh,sorokin},
the most recent of which \cite{sorokin} reports  ionic charge states up to Xe$^{21+}$  at 13.3\,nm and a laser intensity of  about
$10^{16}$\,W/cm$^2$.

The photoeffect becomes highly nonlinear and  requires a nonperturbative description if $K_0=I/\hbar\omega\gg1$, with $I$ the ionization potential.
For intense, low-frequency fields where ionization proceeds via tunneling through the barrier formed by the binding potential plus the quasistatic laser field such nonperturbative descriptions have been developed decades ago.
According to the standard terminology, the tunneling regime is determined by the smallness of the Keldysh parameter \cite{keldysh}, $\gamma=\sqrt{2I}\omega/{\cal E}_0\ll 1$, where ${\cal E}_0$ is the electric field amplitude of the laser wave (we use atomic units unless noted otherwise).
In this limit an analytical expression for the total ionization rate of atoms and positive ions was found for the first time in Ref.~\cite{perelomov67}.
Later this result was extensively generalized, including relativistic ionization and the ionization of molecules or other spatially extended systems (see the review \cite{popov04rev}, also Refs.~\cite{relativistic,brabec}, and references therein).
Compared to tunneling out of a short-range potential, the long-range Coulomb interaction between the outgoing electron and the atomic core of charge ${\cal Z}\ge 1$ strongly increases the tunneling probability
because the Coulomb attraction suppresses the potential barrier through which the photoelectron tunnels.
Experimentally, the Coulomb-induced enhancement of the tunneling ionization rate is well established since the 80s \cite{chin}.
Nowadays the Coulomb-corrected tunneling rates (widely known as ``ADK rates'' \cite{popov66,ADK}) are commonly used for the calibration of laser pulse intensities.
Much less attention has been paid to the theory of the opposite multiphoton limit $\gamma\gg 1$.
For strong fields this regime can only be reached using high frequencies, which is the reason why it became experimentally accessible only with the invention of the high-power FELs.
In the experiments \cite{wabnitz02,laarmann04,wabnitz05,sorokin} typical values of the Keldysh parameter were $\gamma\simeq 30\div 100$.
Tunneling rates are completely irrelevant in this domain: at an intensity $\simeq 10^{16}$\,W/cm$^2$ (as used in Ref.~\cite{sorokin}), for instance, sequential tunneling ionization would predict ions up to Xe$^{8+}$--Xe$^{9+}$ only instead of Xe$^{21+}$, and the probability to ionize Xe$^{10+}$ via tunneling is $\simeq 10^{-24}$ for a 10\,fs laser pulse.

In this Letter an analytical expression for the total (energy and angle-integrated) ionization rate of atoms and positively charged ions in the field of an intense, high-frequency laser pulse, i.e., for $\gamma\gg 1$, is derived, including  the Coulomb-correction, which gives an up to nine orders of magnitude enhancement.
Our rate yields the correct tunneling limit so that it is actually valid at arbitrary frequencies and arbitrary values of the Keldysh parameter. The only limitations are that (i) our rate is restricted to the nonrelativistic regime, and (ii) that only one electron is considered active.
We prove the high accuracy of our result by comparing the predicted rates with {\em ab initio} numerical results.

The ionization rate $w$ of an atom (ion) with the electron in a state of ionization potential $I$ can be represented as the product of the rate $w_{\rm SR}$ for a system bound by a short-range potential with the same ionization potential and the Coulomb correction $Q$, which accounts for the asymptotic Coulomb interaction $U_{\rm C}\simeq-{\cal Z}/r,~r\gg 1/\sqrt{2I}$ between the outgoing electron and the core \cite{faisal01}.
The theoretical approach for the derivation of $Q$ based on the method of classical, complex trajectories (imaginary time method \cite{popov05}) was introduced in \cite{perelomov67} for the tunneling regime $\gamma\ll 1$.
Readers interested in the details of the complex trajectory method are referred to Refs.~\cite{popov05,popov07,poprz08,poprz08jmo,note} where also several applications are presented.
In the semiclassical limit $K_0\gg 1$ the ionization probability is determined by the imaginary part of the reduced classical action evaluated along a trajectory in complex time:
\beq
w\sim\exp\{-2{\rm Im}W\},~~W=\int_{t_s}^{+\infty}\!\!\!\left({\cal L}-I\right)\,\diff t-\vecv\cdot\vecr\bigg\vert_{t_s}^{+\infty}.
\label{wgen}
\eeq
Here, ${\cal L}$ is the Lagrangian, $\vecr(t)$,  $\vecv(t)$ are the electron trajectory and velocity, respectively, and $t_s$ is the complex start time determined by the condition $\vecv^2(t_s)=-2I$.
A trajectory $\vecr(t)$ satisfies Newton's equation of motion.
If only the laser force is accounted for in this equation [we denote a corresponding Coulomb-free trajectory as $\vecr_0(t)$], Eq.~(\ref{wgen}) gives the probability $w_{\rm SR}$ for the ionization from a short-range potential.
This is the well-known output of Keldysh theory \cite{keldysh,popov04rev} or the strong field approximation \cite{faisal}.
A perturbative account of the Coulomb force yields the Coulomb correction
\beq
Q=\exp\{-2{\rm Im}W_{\rm C}\},
\label{Q}
\eeq
\beq
W_{\rm C}=\int_{t_s}^{+\infty}\delta{\cal L}\,\diff t-\delta(\vecv\cdot\vecr)\bigg\vert_{t_s}^{+\infty}+
{\cal Z}\int_{t_s}^{+\infty}\frac{\,\diff t}{r_0(t)}.
\label{WC}
\eeq
Here, $\vecr(t)=\vecr_0(t)+\vecr_1(t)$ is the Coulomb-corrected trajectory, $\delta{\cal L}=\vecv_0\vecv_1+\vecv_1^2/2-{\cal E}(t)\cdot\vecr_1$ is the respective correction to the Lagrangian, $\delta(\vecv\cdot\vecr)=\vecv_0\cdot\vecr_1+\vecv_1\cdot\vecr_0+\vecv_1\cdot\vecr_1$, and ${\cal E}(t)$ is the electric field of the laser pulse.
To eliminate the divergences in (\ref{WC}) a matching with the atomic wave function is required.
This matching technique and other computational details are described in Refs.~\cite{popov07,poprz08jmo}.

In order to find the correction to the total ionization rate it is sufficient to consider the most probable electron trajectory.
In a linearly polarized field ${\cal E}(t)={\cal E}_0\cos(\omega t)$ the  photoelectron momentum vanishes at infinity (i.e., at the detector) for this most probable trajectory.
In the limit $\gamma\gg 1$ the respective Coulomb-free trajectory $x_0(t)$, the correction $x_1(t)$ (both one-dimensional, along the polarization direction) and the complex start time $t_s$ have the form \cite{popov07}
$
x_0(\tau)=b(1-u)$, $u=e^{\tau-\tau_0}$, $0\le\tau\equiv- \imagi \omega t\le\tau_0$,
\beq
x_1(\tau)=b\{ \imagi \sqrt{2\mu}\tau+\mu\left\lbrack\tau^2/2-\ln(1-u)+L_2(u)\right\rbrack\}.
\label{x1}
\eeq
Here $b=\sqrt{2I}/\omega$, $\tau_0=- \imagi \omega t_s\simeq\ln(2\gamma)$, and $L_2(u)$ is the Euler dilogarithm \cite{bateman}.
The parameter $\mu={\cal Z}\omega/(2I)^{3/2}$ determines the relative contribution of the Coulomb field to the trajectory (at $\gamma\gg 1$) so that a perturbative account formally requires, in particular, $\mu\ll 1$ (see the discussion of the applicabily conditions in \cite{poprz08jmo}). Note that because of the term $\sim \sqrt{\mu}$ in (\ref{x1}) we keep the seemingly higher-order term $\vecv_1\cdot\vecr_1$ in $\delta(\vecv\cdot\vecr)$, appearing in (\ref{WC}) (see the corresponding remark in \cite{note}).
After some cumbersome but straightforward algebra one obtains for the Coulomb-correction (\ref{Q})
\beq
Q\simeq\left(\frac{2}{F}\right)^{2n^*}(1+2e^{-1}\gamma)^{-2n^*},~~~~~e=2.718...,
\label{main}
\eeq
where $F={\cal E}_0/(2I)^{3/2}$ is the reduced electric field and $n^*={\cal Z}/\sqrt{2I}$ is the effective principle quantum number of the bound state.
The ionization rate of an atom (ion) then is \cite{faisal01}
\beq
w=Q\cdot w_{\rm SR},
\label{W}
\eeq
where \cite{popov04rev}
$$
w_{\rm SR}=\frac{2C^2}{\pi}IK_0^{-3/2}\beta^{1/2}\sum_{n>n_{\rm th}}{\cal F}(\sqrt{\beta[n-n_{\rm th}]})
\nonumber
$$
\beq
\times\exp\left\{-\frac{2g(\gamma)}{3F}-2c_1(n-n_{\rm th})\right\},
\label{WSR}
\eeq
$$
g(\gamma)=\frac{3}{2\gamma}\left\lbrack\left(1+\frac{1}{2\gamma^2}\right){\rm arcsinh}\,\gamma-\frac{\sqrt{1+\gamma^2}}{2\gamma}\right\rbrack,
\nonumber
$$
$c_1={\rm arcsinh}\,\gamma-\gamma/\sqrt{1+\gamma^2}$, $\beta=2\gamma/\sqrt{1+\gamma^2}$, ${\cal F}(x)$ is the Dawson integral \cite{abram}, and  $n_{\rm th}=K_0[1+1/(2\gamma^2)]$ is the ionization threshold (in units of the photon energy).
For the asymptotic coefficient $C$ of the bound state wave function we use Hartree's approximation \cite{hartree,popov04rev}, $C^2=2^{2n^*-2}/(n^*!)^2$ (for s-states).

Expression (\ref{main}) is our main result.
Although derived for $\gamma\gg 1$, the Coulomb correction (\ref{main}) obeys the right tunneling limit.
Thus for $\gamma\ll 1$ Eq.~(\ref{W}) gives the well-known expression for the tunneling ionization rate of an atomic s-state \cite{perelomov67,popov04rev}.
In the intermediate domain $\gamma\simeq 1$ our result can be considered as an interpolation.
In the limit $\gamma\gg 1$ the correction is intensity-independent and numerically large, $Q\simeq (2eK_0)^{2n^*}\gg 1$.
For noninteger $n_{\rm th}$ the Dawson integral is of the order of unity, the rate (\ref{WSR}) simplifies for $\gamma\gg 1$, and one obtains for (\ref{W})
\begin{eqnarray}
w &\simeq& IA(K_0,n^*)F^{2N_{\rm m}}, \label{Wmult} \\
A(K_0,n^*) &\simeq& 2^{2n^*}C^2e^{N_{\rm m}+2n^*}K_0^{2N_{\rm m}+2n^*-3/2} \nonumber
\end{eqnarray}
where $N_{\rm m}=[K_0]+1$ is the minimum number of photons required for ionization.
The fact that the rate (\ref{Wmult}) is proportional to the intensity to the  power of $N_{\rm m}$ is, of course, not surprising in the multiphoton regime where an atom absorbs the minimum possible number of light quanta \cite{keldysh,popov04rev}.
The new achievement is the derivation of the coefficient $A(K_0,n^*)$ with, as we show below, high quantitative accuracy. Note that the coefficient $A(K_0,n^*)$ is a product of big numbers and thus very sensitive to small variations.
The Coulomb correction (\ref{main}) affects the value of $A$ significantly: for, e.g., the  parameters $K_0\simeq 5\div 10$, $n^*\simeq 1\div 3$, typical for strong-field ionization of atoms and positive ions, $Q$ varies between $10^2$ and $10^{10}$.
The second factor in Eq.(\ref{main}), which distinguishes our Coulomb correction from the tunneling limit, alters the ionization rate also significantly for $\gamma\gg 1$. Note that in Ref.~\cite{faisal01} the Coulomb-free probability $w_{\rm SR}$ was calculated for arbitrary $\gamma$ but the Coulomb correction was taken in the tunneling limit $Q=(2/F)^{2n^*}$.

\begin{figure}
\includegraphics[width=0.38\textwidth]{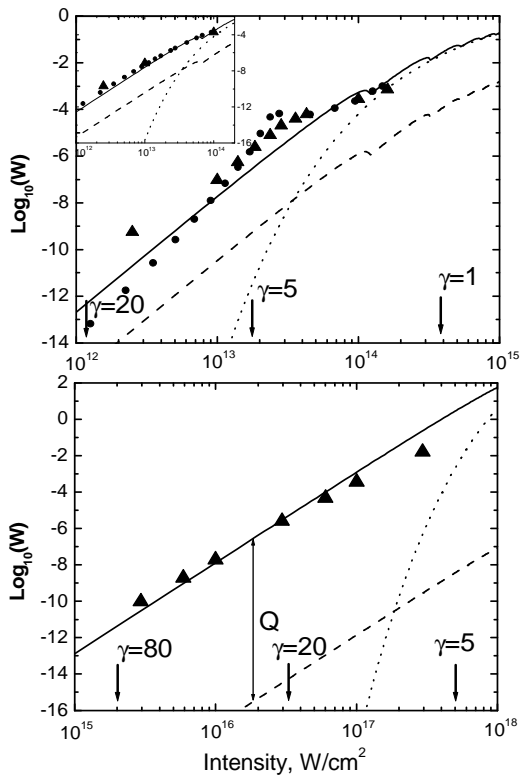}
\caption{Ionization rate (in atomic units) vs laser intensity according (\ref{W}) (solid), (\ref{WSR}) (dashed), and from the respective tunneling formula (dotted line). Rates obtained numerically using the Floquet method and from the solution of the TDSE are shown by dots and triangles, respectively. Upper panel: H(1s) in a field corresponding to the second harmonic of a Ti:Sa laser ($\lambda=400$\,nm). Insert: same but for $\lambda=422$\,nm. Lower panel: Xe$^{17+}$ (4p$_0$-electron with $I=434$\,eV, $n^*=3.19$) in a $\lambda=13.3$-nm x-ray field.}
\label{fig1}
\end{figure}

As examples, the rates (\ref{W}) and (\ref{WSR}) are plotted as a function of  the laser intensity in Fig.~\ref{fig1} for (i) the ionization of atomic hydrogen by the second harmonic of a Ti:Sa laser and (ii) for the ionization of Xe$^{17+}$ by a 13.3-nm x-ray field.
The Coulomb-parameter $\mu$ is 0.11 for H and 0.34 for Xe so that the theory is applicable in both cases. The correction (\ref{main}) is about $6\times 10^2$ and $9\times 10^8$, respectively.
The corresponding tunneling rate is included in both panels, showing that it is completely irrelevant for $\gamma\gg 1$ but merges with (\ref{W}) when approaching the tunneling limit  $\gamma\ll 1$.
In order to check the analytical expressions we compare their predictions with exact numerical results obtained by two different methods: the Floquet method yields the exact ionization rate for an infinite pulse as twice the (absolute value of the) imaginary part of the Floquet-eigenenergy. The solution of the  time-dependent Schr\"odinger equation (TDSE) in real time in general depends on the pulse shape but with increasing pulse duration also yields converged results  for the rate. We used the publicly available codes  STRFLO  \cite{potvliege} and Qprop \cite{koval}. 
In the real-time TDSE calculations the norm ${\cal N}(t)$ inside a sufficiently large test-sphere around the ion is calculated. 
The ionization probability $1-{\cal N}(\infty)$ divided by the  pulse duration $\tau$ is then taken as the ionization rate. Alternatively, one may determine the rate from the slope of ${\cal N}(t)$.
The 4p$_0$ state of Xe$^{17+}$ is constructed as an eigenstate of the effective potential $U(r)=-{\cal Z}/r-(54-{\cal Z})e^{-\kappa r}/r$, where ${\cal Z}=18$ and the parameter $\kappa=7.93$ is adjusted to reproduce the ionization potential $I=434$\,eV.

The comparisons of our analytical rate with the numerical results show, in general, a good, quantitative agreement even for the case of xenon where the parameter $\mu=0.34$ is not so small and the bound state is a p-state while the analytical result (\ref{main})--(\ref{WSR}) is derived for s-states. 
Deviations of our analytical rate from the numerical results can be attributed to resonances.
This is shown in the upper panel  of  Fig.~\ref{fig1} where a 4-photon resonance with a Stark-shifted Rydberg level is met.  This resonant enhancement of the rate at intensities $\simeq 2\div 3\times 10^{13}$\,W/cm$^2$ is more pronounced in the Floquet results because this method assumes an ideally monochromatic pulse.  In the real-time TDSE solution a 10-cycle flat-top pulse with a two cycles up- and down-ramp is used, which has a finite spectral width and thus smears-out resonances.
As it is seen from the insert the analytical result agrees very well with both sets of numerical data when the resonance is avoided by a small variation of the wavelength.

Besides our results in this regime we are only aware of the ones obtained by Santra and Greene for the ionization of Xe at $\hbar\omega=12.7$\,eV and an intensity $10^{13}$\,W/cm$^2$ \cite{santra04}, which were the parameters used in the experiment \cite{wabnitz05} where charge states up to Xe$^{6+}$ were observed.
In Ref.~\cite{santra04} the multiphoton cross-sections $\sigma_N=w_N/j^N$ were found numerically, with $N$ having the same meaning as $N_{\rm m}$ in Eq.\,(\ref{Wmult}) and $j=c{\cal E}_0^2/8\pi\omega$ the photon flux.
The same cross-sections can be easily calculated from Eq.\,(\ref{Wmult}).
A comparison between our results and the cross-sections of Ref.~\cite{santra04} show, in general, reasonable agreement.
For example, we find from Eq.\,(\ref{Wmult}) $\sigma_4(\mathrm{Xe}^{3+})\simeq 2.8\times 10^{-116}$cm$^8$s$^3$ and $\sigma_5(\mathrm{Xe}^{4+})\simeq 5.3\times 10^{-150}$cm$^{10}$s$^4$, while Santra and Greene give $\sigma_4(\mathrm{Xe}^{3+})\simeq 1.8\times 10^{-115}$cm$^8$s$^3$ and $\sigma_5(\mathrm{Xe}^{4+})\simeq 1.1\times 10^{-148}$cm$^{10}$s$^4$, respectively.
Our analytical cross-sections are systematically lower, a fact we attribute to the method used for the extraction of the rate from the numerical data.
More precisely, a complex absorbing potential for distances $r>R_0=4$\,a.u.\ was used in \cite{santra04} so that population of Rydberg states may have contributed to ionization. 

A simple analytical expression for the multiphoton ionization rate for Rydberg states of atoms and ions was derived long ago by Berson \cite{berson} under the condition $n^*\ge N_{\rm m}$ while our case is the opposite, $N_{\rm m}\simeq K_0 > n^*$.
As a consequence, the result of Berson yields a different expression for the coefficient $A(K_0,n^*)$ in  Eq.~(\ref{Wmult}) and does not match the tunneling limit.

In conclusion, we presented an analytical formula for the ionization rate of atoms and ions which is valid for arbitrary values of the Keldysh parameter.
The high accuracy of our rate was demonstrated by comparisons with exact numerical calculations.
In the present form, our theory  discards resonances and describes ionization from s-states in a linearly polarized field.
Generalization to the case of arbitrary polarization and arbitrary angular momentum of the initial state, and the inclusion of relativistic corrections  are straightforward.

Our theory is based on the single active electron approximation. Thus in complex atoms it should describe the sequential ionization channel when electrons are emitted one after another.
It is by now well established, however, that the electron-electron interaction may change the ionization rate dramatically via correlated, nonsequential mechanisms.
In low-frequency fields this nonsequential ionization is essentially understood within the recollision or 'atomic antenna' picture \cite{suran,kuchiev,becker}.
In the high-frequency domain nonsequential mechanisms have been studied both in experiments and theoretically for cases where only a few laser photons are involved (see, e.g. \cite{mosh,istomin,sorokin07,ivanov} and references therein).
In the multiquantum domain we consider here multielectron effects remain to be explored.
The theoretical work by Santra and Greene, e.g., indicates that the cross-sections change up to one order of magnitude if many-electron effects are taken into account \cite{santra04}.
The experiment reported in Ref.~\cite{sorokin} indicates that for highly charged xenon ions the double-logarithmic slope of the intensity-dependent yield is not equal to $N_{\rm m}$ but rather remains constant with increasing charge state.
These facts are hints that multielectron mechanisms may also play an important role in multiquantum ionization at high frequencies.
However, even if this is the case our rate is useful since any measured deviation from the predictions of  Eqs.\,(\ref{main})--(\ref{WSR}) which do not sensitively depend on the wavelength (and therefore must be nonresonant in nature) can then be attributed to electron-electron correlation.

We are grateful to S.P.\ Goreslavski and B.M.\ Karnakov for valuable discussions.
The work was supported by the Deutsche Forschungsgemeinschaft, the Russian Foundation for Basic Research (grants No. 06-02-17370 and 07-02-01116) and by the program of the Russian Ministry of Science and Education for support of the leading research schools (project No.\ 2.1.1.1972).



\begin{thebibliography}{99}
\bibitem{note}  Ref.~\cite{popov07} contains the first attempt to derive the Coulomb correction to the ionization rate for $\gamma\gg 1$, a derivation of the correction (\ref{x1}), and a description of the matching procedure. However, the final result (Eq.\,(14) in \cite{popov07}) is orders of magnitude off because of the erroneous omission of a seemingly higher-order term.
\end{thebibliography}
\end{document}